\documentclass[aps,pra,reprint,showpacs,superscriptaddress]{revtex4-1}
\usepackage[english]{babel}
\usepackage{amsmath,amssymb,bbm,graphicx,color,comment,txfonts}
\usepackage[bookmarks=true,colorlinks,citecolor=blue,urlcolor=blue]{hyperref}
\usepackage{dsfont}
\definecolor{darkgreen}{rgb}{0,0.5,0}
\definecolor{orange}{rgb}{1,0.5,.3}

\usepackage{bbm}
\usepackage{float}

\usepackage{dcolumn}   
\usepackage{bm}        
\usepackage{filecontents}
\usepackage{lineno}

\usepackage{mathtools}
\usepackage{xcolor}

\DeclarePairedDelimiter\ket{\lvert}{\rangle}
\DeclarePairedDelimiterX\braket[2]{\langle}{\rangle}{#1 \delimsize\vert #2}
\newcommand{\mean}[1]{\langle #1 \rangle}

\newcommand{\cre}[2]{{#1}^{\dagger}_{#2}}
\newcommand{\ann}[2]{{#1}^{\phantom{\dagger}}_{#2}}

\newcommand\bs[1]{\ensuremath{\boldsymbol{#1}}}

\begin{document}

\title{Electron spectral functions in a quantum dimer model for topological metals}
\author{Sebastian Huber}
\author{Johannes Feldmeier}
\author{Matthias Punk}
\date{\today}
\affiliation{Physics Department, Arnold Sommerfeld Center for Theoretical Physics and Center for NanoScience, Ludwig-Maximilians-University Munich, 80333 Munich, Germany}

\begin{abstract}
We study single electron spectral functions in a quantum dimer model introduced by Punk, Allais and Sachdev in Ref.~[M. Punk, A. Allais, and S. Sachdev, Proceedings of the National Academy of Sciences 112, 9552 (2015)]. The Hilbert space of this model is spanned by hard-core coverings of the square lattice with two types of dimers: ordinary bosonic spin-singlets, as well as fermionic dimers carrying charge +e and spin 1/2, which can be viewed as bound-states of spinons and holons in a doped resonating valence bond (RVB) liquid. This model realizes a metallic phase with topological order and captures several properties of the pseudogap phase in hole-doped cuprates, such as a reconstructed Fermi surface with small hole-pockets and a highly anisotropic quasiparticle residue in the absence of any broken symmetries.
Using a combination of exact diagonalization and analytical methods we compute electron spectral functions and show that this model indeed exhibits a sizeable antinodal pseudogap, with a momentum dependence deviating from a simple d-wave form, in accordance with experiments on underdoped cuprates.
\end{abstract}

\maketitle

\section{Introduction}\label{Intro}

Since the discovery of high-temperature superconductivity in cuprates \cite{bednorz1986possible}, many efforts have been made to understand the underlying pairing mechanism. A potential solution to this problem might come from a better understanding of the so-called pseudogap state at low doping, out of which the superconducting phase develops upon further doping.  By now we know from a wide range of experiments that the pseudogap state is largely dominated by antiferromagnetic fluctuations \cite{niedermayer1998common,kastner1998magnetic,tallon2001doping,chan2016commensurate} and has a tendency to charge density-wave ordering (CDW) \cite{wu2011magnetic,ghiringhelli2012long,chang2012direct,blanco2013momentum,comin2014symmetry}.

During the last decades experiments have explored the complex physics of the pseudogap phase in much detail.
Early studies of the Knight shift in NMR measurements have shown that the magnetic susceptibility decreases below a characteristic temperature scale, indicative of a spin gap \cite{alloul198989,curro1997high}. Soon measurements of the tunneling density of states \cite{renner1998pseudogap}, c-axis optical conductivity \cite{homes1993optical, uchida1997spin} and specific heat \cite{loram1993electronic,loram2001evidence} revealed that the opening of the pseudogap appears in both, charge and spin degrees of freedom. 

Interestingly, some transport experiments show rather ordinary metallic properties in the pseudogap phase. For example, in-plane
optical conductivity and magnetoresistance measurements show Fermi liquid-like behavior \cite{mirzaei2013spectroscopic,chan2014plane}. 
However, Hall coefficient as well as the Drude weight indicate that the charge carrier density is small and proportional to the density $p$ of doped holes away from half filling \cite{orenstein1990frequency,uchida1991optical,cooper1993optical,padilla2005constant,badoux2015change}, rather than $1+p$ as expected from Luttinger's theorem and observed in the Fermi liquid regime at large doping.
On the other hand, angle resolved photoemission (ARPES) experiments show a distinct Fermi-arc spectrum and a gap opening in the vicinity of the anti-nodes in momentum space \cite{damascelli2003angle,shen2005nodal,hashimoto2015energy,shen1993anomalously,ding1996angle,hussain2007abrupt,vishik2012phase,vishik2010arpes}. Taken together, these observations are hard to reconcile with a Fermi liquid picture, where the Fermi surface is reconstructed by some thermally fluctuating order parameter. In this case one would expect some signatures of the order parameter correlation length to be observable, e.g.~in transport measurements.

A different set of theoretical ideas, based on Anderson's resonating valence bond (RVB) picture \cite{anderson1973resonating,fazekas1974ground,anderson1987resonating}, tries to approach the pseudogap from the Mott insulating state at low doping \cite{lee2006doping}. Upon doping the RVB state with holes, electrons fractionalize into neutral spin-1/2 spinon excitations, as well as spinless holons, carrying charge +e. While electron fractionalization in quasi two-dimensional systems has been a topic of considerable interest on its own, it became clear early that simple incarnations of the RVB state do not capture the sharp Fermi-arc features observed in photoemission experiments, which requires that a certain part of the low-energy spectrum behaves like electron or hole-like quasiparticles which carry both spin and charge.

One possible solution to this puzzle might be provided by the idea of fractionalized Fermi liquids (FL*), introduced originally in the context of heavy Fermion systems \cite{senthil2003fractionalized}. In the context of cuprates, a fractionalized Fermi liquid can be viewed as a doped RVB liquid where spinons and holons form bound states. These hole-like bound states then form a Fermi liquid, with the size of the Fermi surface proportional to the density $p$ of doped holes. Note that the FL* inherits topological order from the RVB background,
which accounts for the violation of Luttinger's theorem \cite{senthil2004weak,kaul2007algebraic,qi2010effective,mei2012luttinger,punk2012fermi,sachdev2016novel}.
The quantum dimer model introduced in Ref.~\cite{punk2015quantum} provides an explicit and intuitive lattice realization of a fractionalized Fermi liquid and allows to directly compute electronic properties. Subsequent numerical work indeed showed clear signatures of a small Fermi surface, indicative of a FL* ground state \cite{lee2016electronic}. In this model bound states between spinons and holons are represented by fermionic dimers on nearest-neighbor sites, which resonate with the background of bosonic spin-singlet dimers. At vanishing doping the model reduces to the well-known Rokhsar-Kivelson model \cite{rokhsar1988superconductivity}. An exact solution has been found along a special line in parameter space for an arbitrary density of fermionic dimers and the ground state can be shown to be an FL* in the vicinity of this line \cite{feldmeier2017exact}. We note that doped quantum dimer models have been studied previously, with doped holes as monomers occupying a single lattice site, carrying no spin \cite{rokhsar1988superconductivity,poilblanc2008properties,lamas2012statistical}. 

In this work our aim is to compute electronic spectral functions for the dimer model introduced in \cite{punk2015quantum}. Using a combination of exact diagonalization and analytical approaches we show that this model exhibits a sizable pseudogap in the antinodal region of the Brillouin zone close to momentum $\mathbf{k}=(0,\pi)$ and symmetry related momenta, in accordance with experimental observations in the pseudogap regime of underdoped cuprates.

This paper is organized as follows: In Sec.~\ref{Model} we give a short overview of the quantum dimer model introduced in \cite{punk2015quantum}, define the single hole spectral function and show how it is related to dimer correlation functions. In Sec.~\ref{Pocket} we discuss our numerical results for ground state properties as well as the spectral functions at zero and at finite temperature. Section \ref{MF} contains an analysis of the spectral functions in terms of a two-mode approximation for the low energy spectrum of the dimer model. Finally, in Sec.~\ref{DIAG} we present a diagrammatic approach to compute the electron dispersion as well as the coherent quasiparticle residuum. We conclude with a discussion in Sec.~\ref{M_S5}.

\section{Dimer model}\label{Model}

\begin{figure}
\centering
 \includegraphics[width=\linewidth]{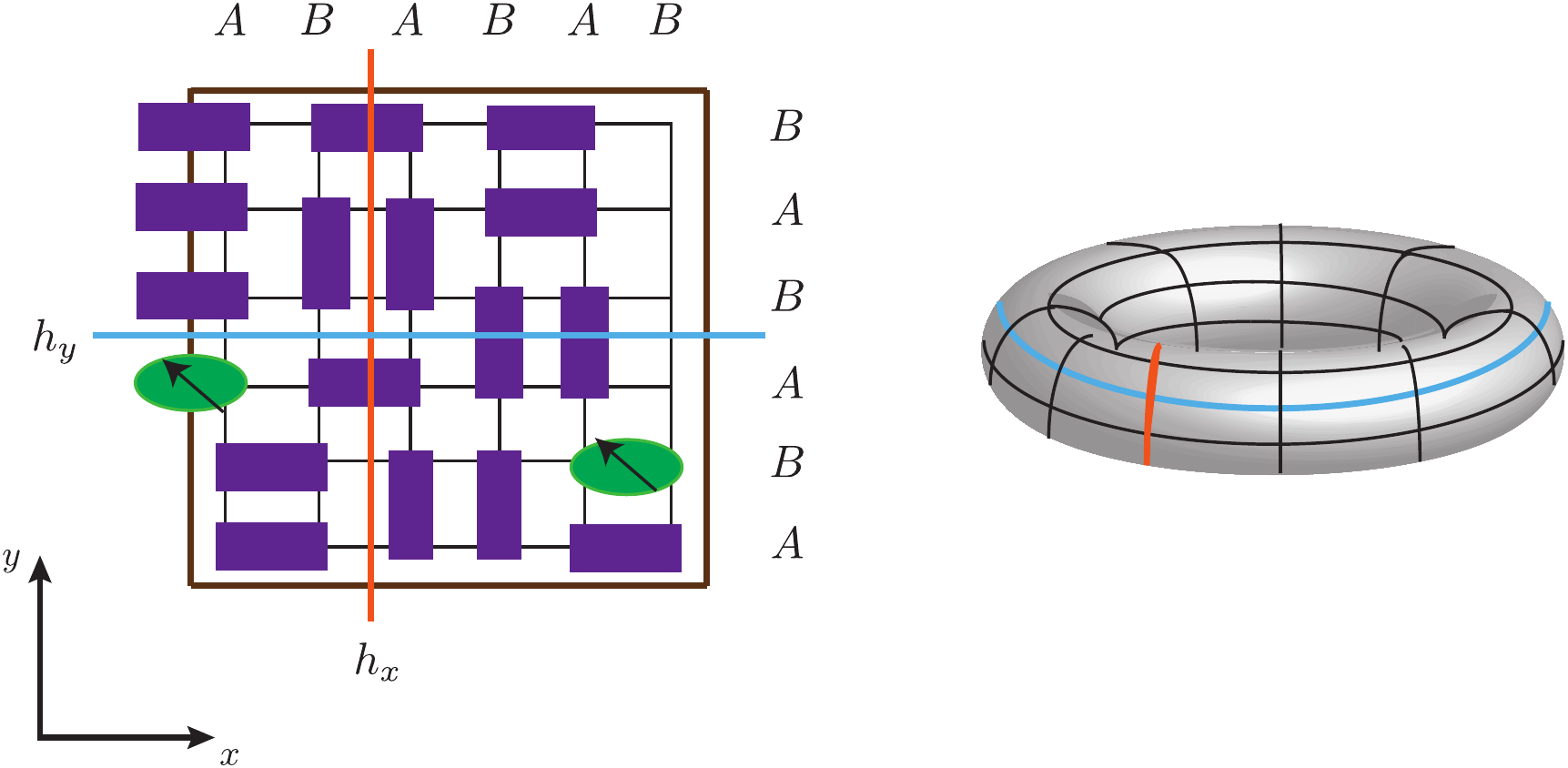}
 \caption{Left: one specific dimer configuration with two fermionic spin S = $1/2$ and charge $+e$ dimers (green ellipses) in a background of bosonic spin singlets (purple rectangles). The blue and orange lines are reference lines to determine the winding numbers of the configuration. Right: Using periodic boundary conditions, the reference lines map onto non-contractible loops on the torus.}
\label{FDimer}
\end{figure}

The Hamiltonian of the quantum dimer model introduced in Ref.~[\onlinecite{punk2015quantum}] acts on a Hilbert space $\mathcal{H_D}= \{ \ket{\mathcal{C}} \}$ spanned by close-packed hard-core configurations $ \ket{\mathcal{C}} $ of two types of dimers living on the links of a two-dimensional square lattice (see Fig.~\ref{FDimer}): the usual bosonic spin-singlet dimers, represented by bosonic operators $\ann{D}{i,\eta}$, as well as fermionic spin-1/2 dimers carrying charge +e, represented by fermionic operators $F_{i,\eta,\alpha}$. Here $i$ denotes the lattice site, $\alpha \in \{\uparrow,\downarrow\}$ is a spin S = $1/2$ index and $\eta \in \{x,y\}$ distinguishes x- and y-links on the square lattice. A fermionic dimer represents a single electron, delocalized in the bonding-orbital between two neighboring lattice sites. Alternatively one can view it as a bound state of a spinon and a holon in a doped RVB liquid.
 
Within the restricted Hilbert-space of the dimer model, the annihilation operator of an electron with spin $\alpha$ on lattice site $i$ can be uniquely expressed in terms of the bosonic and fermionic dimer annihilation and creation operators as \cite{punk2015quantum}\begin{align}
c_{i,\alpha}=\frac{\epsilon_{\alpha\beta}}{2}\sum_{\eta} \left( \cre{F}{i-\hat{\eta},\eta,\beta}\ann{D}{i-\hat{\eta},\eta}+\cre{F}{i,\eta,\beta}\ann{D}{i,\eta} \right) ,\label{MAP}
\end{align}
where $\epsilon_{\alpha\beta}$ is the unit antisymmetric tensor and a sum over repeated spin indices $\beta$ is implied.

We follow Ref.~\cite{punk2015quantum} and consider a Hamiltonian which acts on the states $|\mathcal{C}\rangle$ by resonating dimers of both types along short, flippable loops. The corresponding quantum dimer Hamiltonian reads
\begin{align}
H=H_{RK}+H_{1},\label{HAM}
\end{align}
where 
\begin{align}
H_{RK}=&-J\sum_{i}\cre{D}{i,x}\cre{D}{i+\hat{y},x}\ann{D}{i,y}\ann{D}{i+\hat{x},y} + \text{1 term}\\
&+V\sum_{i}\cre{D}{i,x}\cre{D}{i+\hat{y},x}\ann{D}{i,x}\ann{D}{i+\hat{y},x}  + \text{1 term},\nonumber
\end{align}
is the standard Rokhsar-Kivelson Hamiltonian \cite{rokhsar1988superconductivity} and
\begin{align}
H_{1}=&-t_{1} \sum_{i} \cre{D}{i,x}\cre{F}{i+\hat{y},x,\alpha}\ann{F}{i,x,\alpha}\ann{D}{i+\hat{y},x} + \text{3 terms}\\
&-t_{2}\sum_{i}\cre{D}{i+\hat{x},y}\cre{F}{i,y,\alpha}\ann{F}{i,x,\alpha}\ann{D}{i+\hat{y},x} + \text{7 terms}\nonumber\\
&-t_{3}\sum_{i}\cre{D}{i+\hat{x}+\hat{y},x}\cre{F}{i,y,\alpha}\ann{F}{i+\hat{x}+\hat{y},x,\alpha}\ann{D}{i,y} + \text{7 terms}\nonumber\\
&-t_{3}\sum_{i}\cre{D}{i+2\hat{y},x}\cre{F}{i,y,\alpha}\ann{F}{i+2\hat{y},x,\alpha}\ann{D}{i,y} + \text{7 terms}.\nonumber
\end{align}
describes dimer resonances between a bosonic and a fermionic dimer. Additional terms are related to the ones shown explicitly by lattice symmetries. We note that further terms describing resonances between two fermionic dimers can be included as well, but are not expected to play an important role at low doping, where the density of fermionic dimers is small.

The overlap between two possible dimer configurations can be caluclated using transition graphs and decreases strongly with system size \cite{sutherland1988systems,rokhsar1988superconductivity}. We therefore demand that two different dimer configurations $\ket{\mathcal{C}} \in \mathcal{H_D}$ are orthogonal by construction $\langle \mathcal{C} | \mathcal{C'} \rangle=\delta_{\mathcal{C},\mathcal{C'}}$.

\begin{figure*}
\centering
\includegraphics[width=\textwidth]{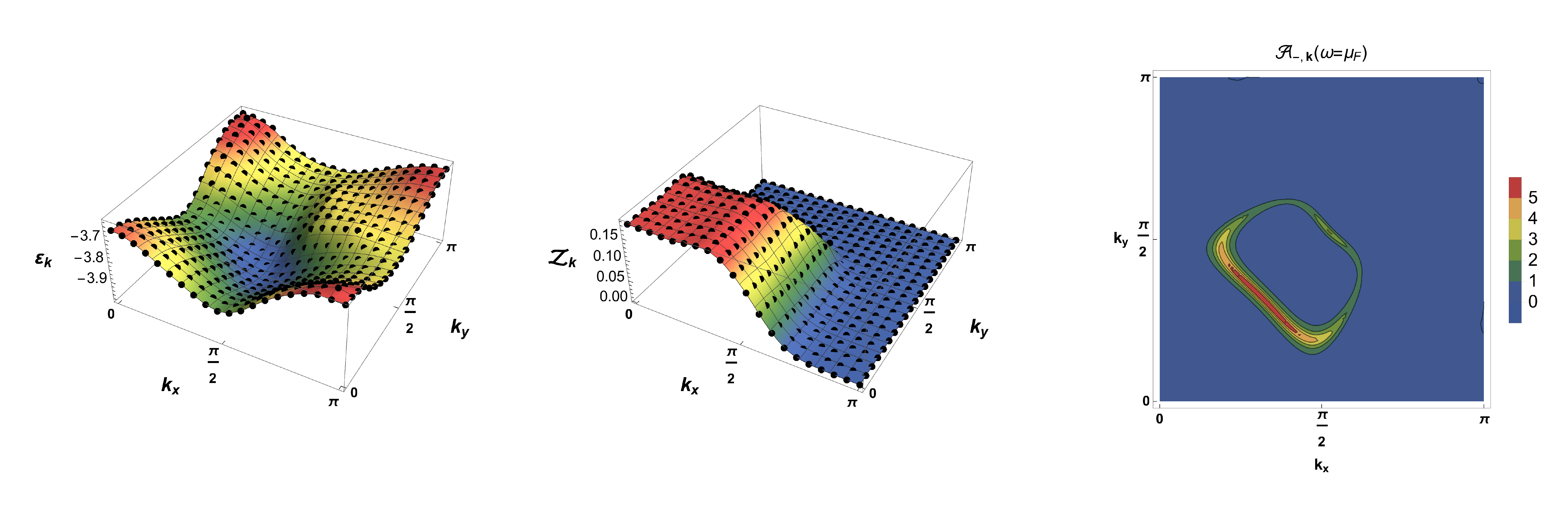}
\caption{Left: energy dispersion $\varepsilon_{\bold{k}}$ of a single fermionic dimer in a background of bosonic dimers, plotted in one quadrant of the Brillouin zone, computed using Lanczos on a $8\times8$ lattice with twisted boundary conditions. Hamiltonian parameters: $J=V=1$, $t_1=-1.05, \, t_2=1.95, \, t_3=-0.6$. Middle: corresponding quasiparticle residue $\mathcal{Z}_\mathbf{k} = |\langle \mathbf{k}| c_{\mathbf{k},\alpha} |0\rangle|^2$. Note the pronounced dispersion minimum in the vicinity of $\bold{k}=\tfrac{1}{2}(\pi,\pi)$, giving rise to pocket Fermi surfaces at a finite density of fermionic dimers, as well as the sharp drop of the quasiparticle residue for momenta larger than $\pi/2$. Right: schematic plot of the coherent part of the spectral function $\mathcal{A}_\mathbf{k}(\omega) \simeq \mathcal{Z}_\mathbf{k} \delta(\omega - \varepsilon_\mathbf{k} + \mu)$ as function of momentum with Lorentzian broadening. Here the chemical potential $\mu$ has been adjusted such that the hole density equals $p\simeq1/8$.}
\label{FPocket}
\end{figure*}

In the following we use periodic boundary conditions (see Fig.~\ref{FDimer}). The Hilbert space $\mathcal{H_D}$ then splits into different topological sectors labeled by a set of two integer winding numbers $\mathcal{W}=\{\mathcal{W}_x,\mathcal{W}_y\}$ defined by 
\begin{align}
\mathcal{W}_{x}=\sum_{i_x }  (-1)^{i_x} \left( \sum_{\alpha} \cre{F}{i,y,\alpha}\ann{F}{i,y,\alpha}+\cre{D}{i,y}\ann{D}{i,y} \right),\\
\mathcal{W}_{y}=\sum_{i_y }  (-1)^{i_y} \left( \sum_{\alpha} \cre{F}{i,x,\alpha}\ann{F}{i,x,\alpha}+\cre{D}{i,x}\ann{D}{i,x} \right).
\end{align}
where the sums run over a line of lattice sites in one direction, counting the staggered number of dimers that cross the reference lines in Fig.~\ref{FDimer}. 
Note that any local Hamiltonian like \eqref{HAM} has non-zero matrix elements only between states in the same topological sector. Matrix elements between states in different topological sectors vanish.

We further make use of symmetries of the quantum dimer Hamiltonian that allow to reduce the computational cost for exact diagonalization. Particle number conservation as well as $SU(2)$ spin-rotation symmetry is implicit in the dimer representation. Finally, the Hamiltonian \eqref{HAM} is invariant under translations as well as square lattice point group symmetries. 
We make use of these symmetries in our numerical implementation to reach a maximum system size of $N=6\times6$ lattice sites for which we calculate the full spectrum for one fermionic dimer embedded in a background of bosonic dimers using exact diagonalization, and a maximum size of $N=8\times8$ sites for the computation of ground-state wavefunctions using a Lanczos algorithm.

In this work our main quantity of interest is the single electron spectral function, which is directly measurable in ARPES experiments \cite{damascelli2003angle,shen2005nodal,hashimoto2015energy,shen1993anomalously,ding1996angle,hussain2007abrupt,vishik2012phase,vishik2010arpes}. The mapping in Eq.~\eqref{MAP} allows us to directly compute the hole-part of the electron spectral function
\begin{align}
\mathcal{A}_{-,\bold{k}}(\omega)&=  \frac{2 \pi}{Z}\sum_{m,n} e^{-\beta E_m} \, |\langle n| c_{\mathbf{k},\alpha} |m\rangle |^2 \delta(\omega - E_m + E_n + \mu)
\label{Ahole}
\end{align}
where $\{ | n \rangle \}$ denotes a complete set of eigenstates of the dimer model with energy $E_n$ and
\begin{align}
c_{\mathbf{k},\alpha} &= \frac{\epsilon_{\alpha\beta}}{2}  \sum_{\mathbf{q},\eta} \left(1+e^{i k_{\eta}} \right) \cre{F}{\mathbf{k}+\mathbf{q},\eta,\beta}  \ann{D}{\mathbf{q},\eta}  
\label{annihilK}
\end{align}

is the electron annihilation operator in the dimer representation. For our numerical computation we fix the particle number and take $|m\rangle$ to be eigenstates of the undoped Rokhsar-Kivelson model, whereas $|n\rangle$ are eigenstates of the dimer model with one fermionic dimer. The full single electron spectral function $\mathcal{A}_{\bold{k}}(\omega) = (1+e^{-\beta \omega}) \, \mathcal{A}_{-,\bold{k}}(\omega)$ follows from the hole-part via detailed balance and is usually normalized as $\int_{\omega} \frac{d\omega}{2\pi}\mathcal{A}_{\bold{k}}(\omega)=1$. It is important to note here that the electron annihilation operator defined in Eq.~\eqref{annihilK} does not obey canonical anticommutation relations, because it is a composite operator. For this reason the positive definite electron spectral function computed here does not satisfy the above normalization condition. The reason for this is that the electron annihilation operator has a non-local representation in the dimer Hilbert space and is dressed by the form factor $f_{\eta} (\bold{k}) = 1+e^{i k_{\eta}}$. The normalization of the spectral function at zero temperature indeed depends explicitly on momentum $\bold{k}$ and is given by $\tfrac{1}{4} [\cos^2{k_x} + \cos^2{k_y}]$, which is thus an upper bound to the quasiparticle residuum shown in Fig.~\ref{FPocket}. The spectrum of the dimer model in our finite size numerics is discrete and the spectral functions are thus composed of a series of delta function peaks with different weight. For better visibility we broaden the delta functions in our figures using a Lorentzian with a width $\delta = 0.04 J$. In order to extract quantitative results for the pseudogap from our numerical spectral function data, we define the gap function $\Delta_{\bold{k}}$ as the distance between lowest energy state at fixed momentum $\mathbf{k}$ and the Fermi energy. It is important to realize that this is a lower bound for the energy gap in the spectral function, because our numerical data shows that the lowest energy states in the spectrum carry a vanishingly small weight in the zero temperature spectral function as one approaches the antinode (see also Fig.~\ref{FPocket} middle), shifting the apparent gap to larger energies. This is no longer true at finite temperatures, however.

\begin{figure*}
\centering
\includegraphics[width=\linewidth]{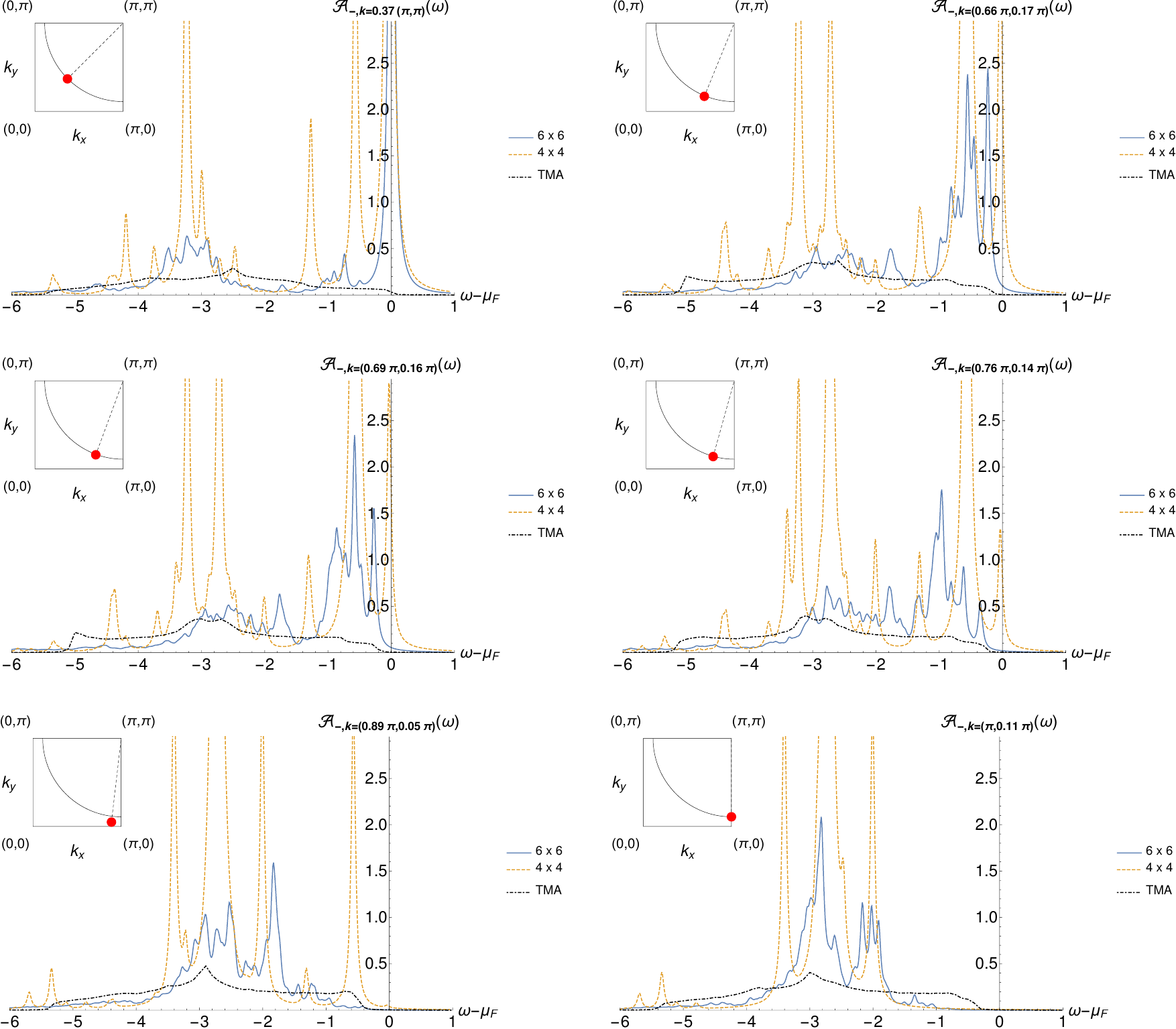} 
\caption{ Top left to bottom right: hole spectral function $\mathcal{A}_{-,\mathbf{k}}(\omega)$ at zero temperature for different momenta $\mathbf{k}$ between the nodal and antinodal region, computed using exact diagonalization for two different system sizes, $4\times4$ (dashed orange line) and $6\times6$ (blue solid line). Parameters:  $J=V=1$, $t_1=-1.05$, $t_2=1.95$, and $t_3=-0.6$. Insets show the respective momentum in the Brillouin zone. The black dash-dotted line corresponds to a two-mode approximation (TMA). The spectral function shows a clear signature of the opening of a pseudogap upon approaching the antinode, independent of system size. At the node (top left) the spectrum exhibits a coherent peak at the Fermi energy, the weight of which is redistributed to the incoherent part of the spectrum at negative frequencies as we move closer to the antinode (bottom right). Note that the TMA is not able to reproduce the coherent peak.}
\label{FSpectral}
\end{figure*} 

\section{Numerical Results}\label{Pocket}

\subsection{Fermi pockets and quasiparticle residue}

The ground state energy $\varepsilon_\mathbf{k}$ of a single fermionic dimer at fixed total momentum $\mathbf{k}$ in a background of bosonic dimers has already been computed in Ref.~\cite{punk2015quantum}, together with the corresponding quasiparticle residue $\mathcal{Z}_\mathbf{k} =  |\langle \mathbf{k}| c_{\mathbf{k},\alpha} |0\rangle|^2$. Here $|\mathbf{k}\rangle$ denotes the ground state of \eqref{HAM} with one fermionic dimer at fixed total momentum $\mathbf{k}$, and $|0\rangle$ is the ground state of the undoped RK-model. In Fig.~\ref{FPocket} we show similar data, but with strongly increased momentum resolution, obtained using a Lanczos algorithm for a $8\times8$ lattice with twisted, rather than periodic boundary conditions, which allows us to compute $\varepsilon_\mathbf{k}$ and $\mathcal{Z}_\mathbf{k}$ for any momentum in the Brillouin zone.

\begin{figure*}
\centering
\includegraphics[width=0.55\linewidth]{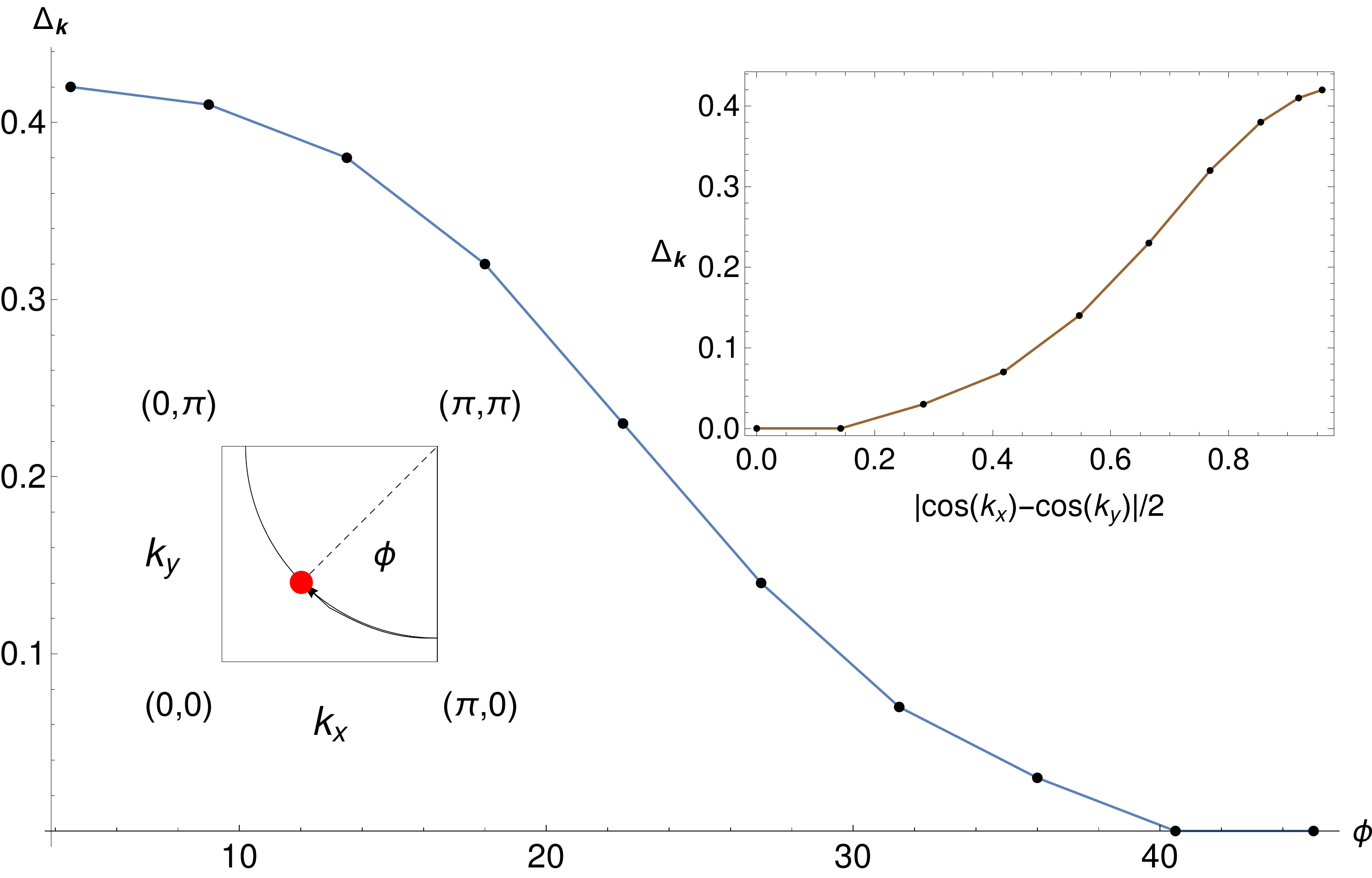}
\caption{Pseudogap $\Delta_\mathbf{k}$, defined as the energy difference between the onset of the spectrum at fixed total momentum $\mathbf{k}$ and the chemical potential, plotted as function of the angle between the antinode and the node (see bottom left inset for a definition of the angle) for a $6\times6$ lattice. Hamiltonian parameters are identical to Fig.~\ref{FSpectral}.
The top right inset shows the clear deviation of $\Delta_\mathbf{k}$ from a pure d-wave form factor $\sim \cos k_x -\cos k_y$. }
\label{Gap}
\end{figure*}

Our results for parameters $J=V=1$, $t_1=-1.05$, $t_2=1.95$, and $t_3=-0.6$, which follow from electron hopping parameters in an effective $t-J$ model appropriate for cuprates (see Ref.~\cite{punk2015quantum} for details), show a pronounced dispersion minimum in the vicinity, but not directly at momentum $\mathbf{k}=(\pi/2,\pi/2)$. We note that the position of the dispersion minimum depends on microscopic details and the precise value of the $t_i$ amplitudes. At a finite density of fermionic dimers this would give rise to a Fermi surface comprised of small hole-pockets with an area proportional to the density of fermionic dimers, which equals the density $p$ of doped holes away from half filling, realizing a fractionalized Fermi liquid. The corresponding quasiparticle residue $\mathcal{Z}_\mathbf{k}$ drops sharply for momenta larger than $\pi/2$ and the associated photoemission response, which is proportional to the hole-part of the electron spectral function, indeed shows Fermi-arc like features due to the highly anisotropic quasiparticle residue. Note that the ground state at a finite density of fermonic dimers potentially breaks symmetries in certain parameter regimes and is not necessarily a fractionalized Fermi liquid with a Fermi surface encompassing an area proportional to the density of fermionic dimers. While the precise nature of the ground state is not yet known in the full parameter space, numerical studies showed the presence of Friedel oscillations associated with a small Fermi surface in accordance with an FL* ground state \cite{lee2016electronic}. Moreover, a recent exact solution of the dimer model also shows that the ground state is an FL* in an interesting parameter regime \cite{feldmeier2017exact}.

\subsection{Spectral functions and pseudogap}\label{PG1}

In Fig.~\ref{FSpectral} we display results for the hole spectral function Eq.~\eqref{Ahole} at zero temperature along a series of momenta between the nodal point on the Fermi pocket and the antinode at $\bold{k}=(\pi,0)$ for two different system sizes, $4\times4$ and $6\times6$ lattice sites, where we use the same parameters in the Hamiltonian as in the previous section ($J=V=1$, $t_1=-1.05$, $t_2=1.95$, and $t_3=-0.6$). At zero temperature we only have to consider matrix elements between the RVB ground state of the Rokhsar-Kivelson model and eigenstates of \eqref{HAM} with one fermionic dimer that belong to the zero winding number sector $(\mathcal{W}_{x},\mathcal{W}_{y})=(0,0)$. At $J=V=1$ the RVB ground state 
$\ket{\Psi}_{\text{RVB}}=\sum_{\mathcal{C}}\ket{\mathcal{C}}$
is an equal amplitude superposition of all dimer configurations and has vanishing energy. 
Again, we used twisted boundary conditions to access arbitrary momenta within the first Brillouin zone.

The top left panel shows $\mathcal{A}_{-,\bold{k}}(\omega)$ at the nodal point, which has a sharp peak at the Fermi surface ($\omega=\mu_F$) independent of system size, as expected from our ground state computations. The incoherent part of the spectrum, containing $\sim 40 \%$ of the hole spectral weight, is broadly distributed with a maximum located around $\omega-\mu_F \approx -3.1J$.
The $6\times6$ data clearly shows that the spectral weight of the central peak vanishes and a gap opens as we go away from the nodal point towards the antinode, where the spectral weight is redistributed to lower frequencies. In the antinodal region around $\mathbf{k}=(\pi,0)$, the spectral function exhibits a sizable pseudogap on the order of $J$, independent of system size (see Fig.~\ref{FSpectral} bottom right).
 The complete spectral weight is now distributed over a broad peak of width $\sim3J$ centered around $\omega-\mu_F \approx -2.8J$. 

In Fig.~\ref{Gap} we plot the pseudogap $\Delta_\mathbf{k}$ as function of the angle $\phi$ between the antinode ($\phi=0$) and the node ($\phi=45^\circ$) for a fixed distance from $\mathbf{k}=(\pi,\pi)$, extracted from our numerical data for a lattice size of $6\times6$ sites. We emphasize again that $\Delta_\mathbf{k}$ corresponds to the onset of the spectral function at low energies and represents a lower bound for the pseudogap extracted from the spectral function, because the low energy states in the spectrum turn out to have a vanishingly small weight in the antinodal region, which leads to an apparently larger gap in the spectral function. This effect is only seen at zero temperature, however. Nonetheless, the dimer model features a sizeable pseudogap on the order of $\Delta_\mathbf{k} \sim 0.4$ at the antinode. It is also important to note that the pseudogap shows a clear deviation from a simple d-wave form $\Delta_\mathbf{k} \sim \cos k_x -\cos k_y$ (see inset in Fig.~\ref{Gap}) and is thus clearly distinguishable from the superconducting gap. This is in agreement with a wide range of experiments that found evidence for corrections to the d-wave symmetry for the pseudogap in the underdoped regime, in stark contrast to the superconducting gap \cite{hussain2007abrupt,vishik2010arpes,loret2017vertical}.

\begin{figure}
\centering
\includegraphics[width=\columnwidth]{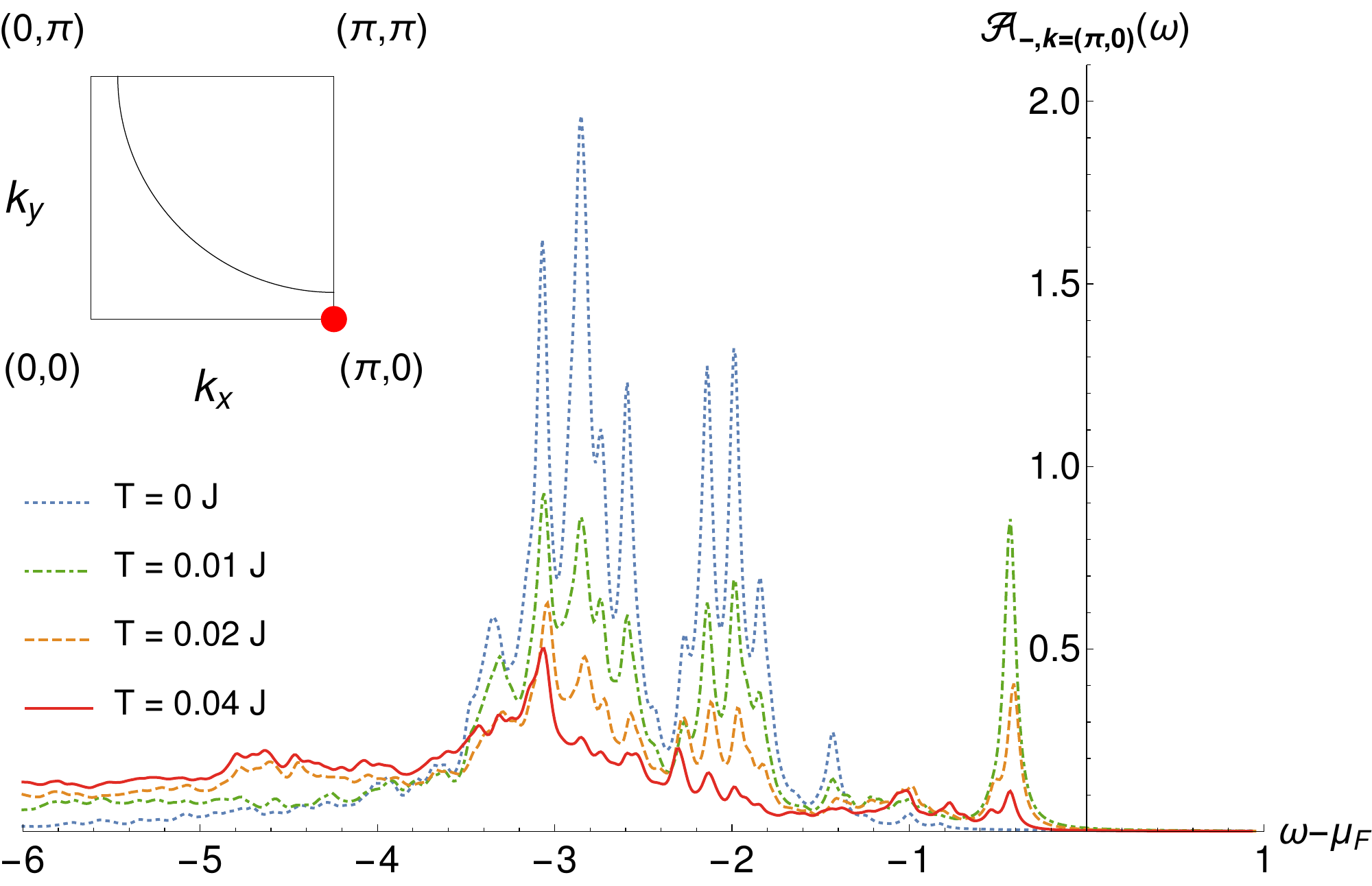}
\caption{Hole spectral function $\mathcal{A}_{-,\mathbf{k}}(\omega)$ at the antinode $\mathbf{k}=(0,\pi)$ for four different temperatures $T=0, \, 0.01J, \, 0.02J, \, 0.04J$. The Hamiltonian parameters are the same as in Fig.~\ref{FSpectral} and only numerical data for $6\times6$ lattices is shown. At finite temperature the gap corresponds to $\Delta_\mathbf{k}$ plotted in Fig.~\ref{Gap}, whereas at zero temperature the gap is apparently larger because the low energy states have a vanishingly small weight.}
\label{SF_T}
\end{figure}

Finally, in Fig.~\ref{SF_T} we show finite temperature results for the hole spectral function in the antinodal region. At finite temperature matrix elements involving excited states of the Rokshar-Kivelson model contribute to the spectrum as well, which leads to a thermal broadening of the incoherent peak and a shift of spectral weight to lower energies.
More interestingly, the finite temperature results show that the low energy states do contribute finite weight to the spectral function at the antinode, and the pseudogap is indeed given by $\Delta_\mathbf{k}$ shown in Fig.~\ref{Gap}.
This is in contrast to the zero temperature case, where low energy states have a vanishingly small weight and the pseudogap appears to be larger in the spectral function.

\section{Two-mode approximation}\label{MF}

In this section we present a semi-analytic description of the hole spectral function in terms of a two-mode approximation, where we assume that the low energy part of the spectrum of excited states can be captured by states of the form
\begin{align}
\ket{p}_k &=\cre{\tilde{F}}{k+p,1}\ann{\tilde{D}}{p,1}\ket{\text{GS}}  \label{ApproxExcit}
\end{align}
where the fermionic (bosonic) operator $\tilde{F}_{k,1}$ ($\tilde{D}_{k,1}$) is a linear combination of $F_{k,x}$ ($D_{k,x}$) and $F_{k,y}$ ($D_{k,y}$) and  describes the lower of the two fermionic (bosonic) bands. $|\text{GS}\rangle$ denotes the ground state of \eqref{HAM} at a given doping.  
In order to determine the matrix elements relating $\tilde{F}_{k,1}$ and $F_{k,\eta}$ we use a mean-field approximation to diagonalize the Hamiltonian \eqref{HAM}, as outlined below. A similar mean-field approach has been used previously in Ref.~\cite{goldstein2017d}.

It is important to note here that the two-mode approximation captures the incoherent part of the hole spectral function, but is not able to reproduce the coherent peak, as we explain in detail below. Moreover, the mean-field approach violates the hard-core constraint for the dimers, which is a very crude approximation. Nevertheless, the main features of the pseudogap are correctly reproduced when approaching the antinodal region in momentum space.

We start by outlining the mean field approximation for the description of a small, but finite density of fermionic dimers interacting with a background of bosonic dimers forming an RVB spin liquid. In total we need five real, homogeneous and isotropic fields to decouple the dimer Hamiltonian in all possible bosonic channels
\begin{align}
\chi^{s}_{z} = \mean{\cre{D}{i,\eta}\ann{D}{j,\eta'}}, \label{MeanFL} 
\end{align}
where $s=0$ denotes on-site terms ($i=j$), whereas $s=1$ are nearest-neighbor terms ($j=i\pm\hat{x},\hat{y}$). The index $z=||, \perp$ denotes parallel ($\hat{\eta}\ ||\ \hat{\eta}'$) or perpendicular ($\hat{\eta} \perp \hat{\eta}'$) dimer configurations.\\
We demand that the mean-field solutions represent a state with no broken symmetries, thus $\chi^{s}_{z}$ must be invariant under the symmetry group of the square lattice. From this follows that the mean-fields $\chi^{1}_{\perp}$ and $\chi^{0}_{\perp}$ have to be equal. The RK mean-field Hamiltonian thus takes the form
\begin{widetext}
\begin{align}
H^{\text{MF}}_{\text{RK}}=\sum_{q} (\cre{D}{q,x},\cre{D}{q,y}) \begin{pmatrix}
V\chi^{0}_{||}+V\chi^{1}_{||} \cos(q_{y}) & -\tfrac{J}{2}\chi^{0}_{\perp} (1+e^{i (q_x-q_y)})-\tfrac{J}{2}\chi^{1}_{\perp} (e^{-i q_y}+e^{i q_x}) \\
-\tfrac{J}{2}\chi^{0}_{\perp} (1+e^{-i (q_x-q_y)})-\tfrac{J}{2}\chi^{1}_{\perp} (e^{i q_y}+e^{-i q_x}) & V \chi^{0}_{||}+V\chi^{1}_{||} \cos(q_{x}) 
\end{pmatrix}
\begin{pmatrix}
\ann{D}{q,x}\\
\ann{D}{q,y}
\end{pmatrix}.
\end{align}
\end{widetext}
Similarly, the interaction term $H_1$ between fermionic and bosonic dimers can be decoupled using the mean-fields \eqref{MeanFL} and reads
\begin{widetext}
\begin{align}
H^{\text{MF}}_{1}=\sum_{q} (\cre{F}{q,x},\cre{F}{q,y}) \begin{pmatrix}
-2t_1\chi^{1}_{||} \cos(q_{y})  & -t_2 \chi^{1}_{\perp} h_{2,q} - t_3 \chi^{2}_{\perp} h_{3,q} \\
-t_2 \chi^{1}_{\perp} \bar{h}_{2,q} - t_3 \chi^{2}_{\perp} \bar{h}_{3,q} & -2t_1\chi^{1}_{||} \cos(q_{x})
\end{pmatrix}
\begin{pmatrix}
\ann{F}{q,x}\\
\ann{F}{q,y}
\end{pmatrix},
\end{align}
with
\begin{align}
h_{2,q} &= (1 + e^{i q_x} + e^{-i q_y} + e^{i(q_x - q_y)}),\\
h_{3,q} &= (e^{i (q_y - 2q_x)} + e^{i q_x} + e^{-i 2q_x} + e^{i (q_x+q_y)} + e^{i2q_y} + e^{i(2q_y-q_x)} + e^{-iq_y} + e^{-i(q_x+q_y)} ).
\end{align}
\end{widetext}
We diagonalize the mean-field Hamiltonian using a Bogoliubov transformation $(\ann{D}{q,x},\ann{D}{q,y})^T=M_q\ (\ann{\tilde{D}}{q,1},\ann{\tilde{D}}{q,2})^T$  and $(\ann{F}{q,x},\ann{F}{q,y})^T =  N_q\ (\ann{\tilde{F}}{q,1},\ann{\tilde{F}}{q,2})^T$, which provides the matrix elements relating $\tilde{F}$ and $F$, as mentioned above. The single particle operators $\ann{\tilde{F}}{q,\alpha}$ and $\ann{\tilde{D}}{q,\alpha}$ describe excitations with dispersion relations $\xi^{\tilde{F}}_{q,\alpha}$, respectively $\xi^{\tilde{D}}_{q,\alpha}$, with band index $\alpha \in \{ 1,2 \}$.

Assuming that the electronic spectrum is well approximated by excitations of the form \eqref{ApproxExcit}, the corresponding eigenenergies of the excitations are then given by
\begin{align}
H^{\text{MF}}\ket{p}_k&=[\xi^{\tilde{F}}_{k+p,1}-\xi^{\tilde{D}}_{p,1}+\xi_{\text{GS}}]\ket{p}_k,\nonumber
\end{align}
where $H^{\text{MF}}=H^{\text{MF}}_{\text{RK}}+H^{\text{MF}}_{1}$ and $H^{\text{MF}}\ket{\text{GS}}=\xi_{\text{GS}}\ket{\text{GS}}$.

The hole part of the spectral function at zero temperature in this mean-field approximation reads
\begin{align}
\mathcal{A}_{-,\bold{k}}(\omega) &=\sum_{p} Q_{\bold{k}}(\bold{p})\ \delta(\omega + \xi^{\tilde{F}}_{k+p,1} - \xi^{\tilde{D}}_{p,1}+\xi_{\text{GS}}),  \label{MF_SF}
\end{align}
where 
\begin{align}
Q_{\bold{k}}(\bold{p})&=[f_k(p) (1-n_{F}(\xi^{\tilde{F}}_{k+p,1}))n_{B}(\xi^{\tilde{D}}_{p,1})]^2,\nonumber\\
f_{k}(\bold{p})&=\tfrac{1}{\sqrt{2N_xN_y}}[\bar{N}^{11}_{p+k} M^{11}_p f_{\hat{x}}(k) + \bar{N}^{21}_{p+k} M^{21}_p f_{\hat{y}}(k)].\nonumber
\end{align}
Note that the spectral function in the two-mode approximation cannot have a sharp, coherent delta-function peak, because Eq.~\eqref{MF_SF} always involves an integral over momenta. 
We use the numerical data of the dispersion $\varepsilon_k$ in Fig.~\ref{FPocket} to constrain the fitting of the mean field dispersions $\xi^{\tilde{F}}_{k,1}$ and $\xi^{\tilde{D}}_{k,1}$ by using the relation $\varepsilon_{k}=\text{min}_p [\xi^{\tilde{F}}_{k+p,1} - \xi^{\tilde{D}}_{p,1}]$, which determines the onset of the spectrum. 
Furthermore we approximate the momentum distribution of fermionic dimers $n_{F}(\xi^{\tilde{F}}_{k+p,1})$ by a simple Fermi-Dirac distribution, which is appropriate within the mean-field description. For the bosonic dimers a Bose-Einstein distribution would not be appropriate, however, because it does not capture the important hard-core constraint of bosonic dimers. For this reason we take the bosonic dimer distribution to be a constant, as expected for hard core bosons in a semi-classcial limit \cite{coletta2012semiclassical}.

In Fig.~\ref{FSpectral} we plot the spectral functions together with the two-mode approximation for different momenta.
Even though this appraoch is not capable to reproduce the coherent peak of the spectral function, we find a finite weight at the chemical potential. Upon approaching the antinodal region the pseudogap slowly opens as function of momentum and has a similar size as in the ED results (see Fig. \ref{FSpectral} bottom left).
The semi-analytic mean field approach describes the spectral function at the antinode reasonably well, where it shows a clear siginature of the pseudogap with a dominant incoherent peak at $\omega-\mu_F \approx -3J$ (see Fig. \ref{FSpectral} bottom right).

In conclusion, the two-mode approximation is able to repoduce some common freatures of the spectral function. Its drawback is that it doesn't capture the coherent quasiparticle peak, however. Such a coherent peak would appear in the two-mode approximation only if we allow for a boson condensate. Since there is no physical basis for the appearance of a condensate in a hard-core boson system at integer filling, we refrained from such a modification.  Another modification would be the inclusion of corrections to the boson distribution $n_B$ beyond the semiclassical limit, as discussed in Ref.~\cite{coletta2012semiclassical}. These corrections give rise to a term $\sim1/p$ in the momentum distribution of the bosonic dimers which leads to the appearance of an additional peak in the spectral function at high energies around $\omega-\mu_F \approx -6J$, but don't change the qualitative behavior at lower energies, in particular the onset of the pseudogap.

\section{Diagrammatic Results}\label{DIAG}

In this section we present a systematic diagrammatic approach to compute the electron spectral function and the coherent quasiparticle residuum in particular. Since the Hamiltonian of the dimer model does not feature a quadratic part, there formally does not exist a small parameter in the system that would rigorously justify the use of such perturbative means. We pursue this approach nonetheless and for $|t_i|\ll J=1$ find good agreement with numerical results. The essential ingredient is the expansion around the exactly solvable $t_i=0$ RK-point \cite{rokhsar1988superconductivity}, where dimer-dimer correlations take a classical form, see e.g.~\cite{fisher1963statistical}.

Within the domain $|t_i|\ll J$, fermionic dimers are inserted into a RK-like bosonic background which most importantly allows to consider the ground state to be translationally invariant. We expect the coherent part of the spectral function to be induced by the part $H_1$ of the dimer Hamiltonian  \eqref{HAM} in which we will conduct a perturbative expansion starting from the action $S$ of the model in momentum space
\begin{align}
S= &\sum_{q_1,q_2,q_3,q_4} \biggl\{ H\left[\bar{F}(q_1),F(q_4),\bar{D}(q_2),D(q_3)\right]\biggr\}+ \label{eq:diag1} \\
&+ \sum_{q,\eta,\alpha}\bar{F}_{\bs{q},\eta,\alpha}(i\omega)\left(-i\omega-\mu_f\right)F_{\bs{q},\eta,\alpha}(i\omega)+ \nonumber\\
&+ \sum_{q,\eta}\bar{D}_{\bs{q},\eta}(i\omega)\left(-i\omega-\mu_b\right)D_{\bs{q},\eta}(i\omega),\nonumber
\end{align}
where in the notation $q\equiv (i\omega_q, \bs{q})$ momentum and energy conservation $q_1+q_2=q_3+q_4$ is understood. The fields $\bar{F}$, $F$ correspond to anticommuting Grassmann variables while $\bar{D}$, $D$ are complex fields. Since $SU(2)$ symmetry is manifest, we drop the spin index for the fermionic fields in the following. As a further approximation we do not enforce the hard-core dimer constraint exactly for the moment, but introduce chemical potentials $\mu_f$ and $\mu_b$ to fix the fermionic and bosonic dimer densities on average (we comment on how to enforce the hard-core constraint below Eq.~\eqref{eq:diag10}). Since the individual terms in the Hamiltonian obey the hard-core constraint locally, this turns out to be a good approximation. The bare dimer propagators thus are 
\begin{align}
G_{f/b}^0(i\omega)=\frac{1}{i\omega+\mu_{f/b}}, \label{eq:diag2}
\end{align}
with doping dependent chemical potentials fixing the average dimer densities on a given link of the lattice to $n_F(-\mu_f)=\frac{p}{4}$ and $n_B(-\mu_b)=\frac{1-p}{4}$. Within the Matsubara formalism we can compute the electronic spectral function $\mathcal{A}_{\bs{p}}(\omega_p)=-2\Im [G^R_{\bs{p}}(\omega_p)]$ from the electronic imaginary time ordered Green's function via analytic continuation $G^R_{\bs{p}}(\omega_p)=\mathcal{G}_{\bs{p}}(i\omega_p\rightarrow \omega_p+i0^+)$. In the dimer Hilbert space (see Eq. \eqref{MAP}), $\mathcal{G}$ can be obtained from the relation
\begin{align}
\mathcal{G}_{\bs{p}}(i\omega_p)=& \frac{1}{4\beta N}\sum_{\eta_1,\eta_2\epsilon\{x,y\}}\left(1+e^{-ip_{\eta_1}}\right)\left(1+e^{ip_{\eta_2}}\right) \label{eq:diag3}\\
\times & \sum_{q_1,q_2}\Big\langle F_{q_2,\eta_2}D_{q_1+p,\eta_1}\bar{D}_{q_2+p,\eta_2}\bar{F}_{q_1,\eta_1}\Big\rangle, \nonumber
\end{align}
where the average $\Big\langle F_{q_2,\eta_2}D_{q_1+p,\eta_1}\bar{D}_{q_2+p,\eta_2}\bar{F}_{q_1,\eta_1}\Big\rangle$ is to be evaluated in the framework of the action $S$ from Eq. \eqref{eq:diag1} ($N$ is the number of sites). The corresponding spectral function for holes is then given by $\mathcal{A}_{-\bs{p}}(-\omega)$.

We can write the four point dimer correlator in full generality as
\begin{widetext}
\begin{align}
\Big\langle F_{q_2,\eta_2}D_{q_1+p,\eta_1}\bar{D}_{q_2+p,\eta_2}\bar{F}_{q_1,\eta_1}\Big\rangle= & \delta_{q_1,q_2}\delta_{\eta_1,\eta_2}G_f(q_1)G_b(q_1+p)+ \label{eq:diag4}\\
+&G_f(q_2)G_b(q_1+p)\, \tilde{\Gamma}^{\eta_1,\eta_2,\eta_1,\eta_2}(q_1,q_2+p,q_1+p,q_2)\, G_b(q_2+p)G_f(q_1), \nonumber
\end{align}
\end{widetext}
where we have introduced the full interaction vertex $\tilde{\Gamma}^{\eta_1,\eta_2,\eta_1,\eta_2}(q_1,q_2+p,q_1+p,q_2)$ and the full (dressed) propagators $G_f$ and $G_b$. It is important to realize that the full propagators can not develop a dispersion at any order in perturbation theory, as the interaction terms in the Hamiltonian locally respect the hard-core constraint. Any diagram with only two external lines thus necessarily has propagators on the same lattice site for the incoming and outgoing lines. Typically, we expect the full propagators to contain finite lifetimes for the two dimer species which merely lead to a Lorentzian broadening of delta-type peaks in the spectral function. We thus approximate $G_{f/b}\rightarrow G_{f/b}^0$ in Eq. \eqref{eq:diag4}. The first term in Eq.~\eqref{eq:diag4} corresponds to the zeroth order contribution and upon insertion into Eq.~\eqref{eq:diag3} results in the electronic spectral function
\begin{align} 
\mathcal{A}_{\bs{p}}^0(\omega_p)=\frac{1}{4}\left(\cos^2 \left(\frac{p_x}{2}\right)+\cos^2\left(\frac{p_y}{2}\right)\right)\cdot 2\pi\delta (\omega_p) \label{eq:diag5}
\end{align}
for the non-interaction limit $t_i=0$. Here, $\mu_{f/b}\xrightarrow{T\rightarrow 0}0$, which leads to the peak position at $\omega_p=0$, was used. It can be verified that this expression indeed reproduces the exact quasiparticle residuum at this specific point in parameter space. We have thus shown that our expansion around an RK-like background at $t_i=0$ reproduces the correct result right at the expansion point as required for a meaningful Ansatz.

\begin{figure*}
\includegraphics[trim={0 0cm 0 0cm},clip,width=\linewidth]{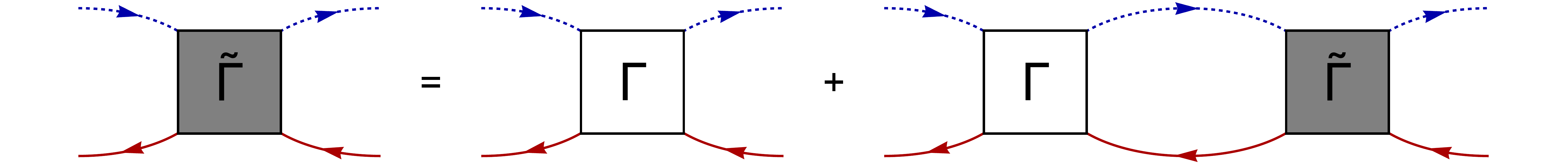}
\caption{Bethe-Salpeter equation for the effective 2-particle vertex $\tilde{\Gamma}$. Dashed blue lines correspond to bare bosonic, solid red lines to bare fermionic propagators. The approach effectively sums up all particle-hole like ladder diagrams.}
\label{fig:bethesalpeter}
\end{figure*}

We now seek to find a good approximation to the full vertex by considering particle-hole like ladder diagrams. These turn out to correspond to repeated exchange hoppings of a fermionic dimer in the bosonic background and are thus expected to yield good estimates for considering exchange terms only. The effective vertex is then determined by solving a Bethe-Salpter equation which is displayed diagrammatically in Fig.~\ref{fig:bethesalpeter}. For the vertex from Eq.~\eqref{eq:diag4} this equation reads
\begin{widetext}
\begin{align}
 \tilde{\Gamma}^{\eta_1\eta_2\eta_1\eta_2}(q_1,q_2+p,q_1+p,q_2)=& \Gamma^{\eta_1\eta_2\eta_1\eta_2}(q_1,q_2+p,q_1+p,q_2)+ \label{eq:diag7}\\
& +\sum_{\tilde{\eta}_f,\tilde{\eta}_b}\beta\psi(i\omega_p)\biggl\{\sum_{\bs{\tilde{q}}}\Gamma^{\eta_1\tilde{\eta}_b\eta_1\tilde{\eta}_f}(q_1,\tilde{q}+p,q_1+p,\tilde{q})\;\;\tilde{\Gamma}^{\tilde{\eta}_f\eta_2\tilde{\eta}_b\eta_2}(\tilde{q},q_2+p,\tilde{q}+p,q_2)\biggr\}, \nonumber
\end{align}
\end{widetext}
where $\Gamma$ corresponds to the bare vertex (see Appendix) which is derived from the Hamiltonian, while the particle-hole bubble $\psi (i\omega_p)=\frac{1}{4}\frac{1}{i\omega_p+\mu_b-\mu_f}$ corresponds to the Matsubara sum of an antiparallel pair of bare fermionic and bosonic propagators.

To solve this integral equation for $\tilde{\Gamma}$, we note that in principle only the exchange interactions $t_1$ and $t_3$ contribute to the ladder diagrams of Eq. \eqref{eq:diag7}. This is due to the external dimer orientations in Eq. \eqref{eq:diag7} being fixed such that the orientation of the ingoing (outgoing) fermionic (bosonic) dimer $\eta_2$ ($\eta_1$) matches the orientation of the outgoing (ingoing) bosonic (fermionic) dimer. A $t_2$ flip term would induce a different relative orientation between ingoing and outgoing dimers and thus cannot contribute to the particle-hole ladder. However, we can include the $t_2$-terms in an effective manner by substituting them with the following exchange terms,
\begin{align}
H_{t_2}\rightarrow \tilde{H}_{t_2}=-t_2\sum_{i}\cre{D}{i,y}\cre{F}{i+\hat{y},x,\alpha}\ann{F}{i,y,\alpha}\ann{D}{i+\hat{y},x} + \text{7 terms}.\label{eq:diag14}
\end{align}
Even though this term explicitly violates the hard-core constraint, the physical reasoning for this replacement is the following: we expect the major contributions by $t_2$-processes to come via induced exchange interactions, i.e. a $t_2$-flip combines with a $t_1$- or $t_3$- exchange and a subsequent bosonic $J$-plaquette flip which restores the starting configuration of the bosonic background. Such a sequence effectively acts as an exchange. Since our ladder approach does not include the purely bosonic background interactions, we account for these by assigning to $t_2$ an effective first order exchange interaction that induces the correct exchange interactions at higher order. Note that although the terms in Eq. \eqref{eq:diag14} project on non-constraint configurations, we still expect this substitution to be valid as we consider a translational invariant dimer density in our approach. Comparing our diagrammatic results to numerical data shows that this approximation indeed works very well.

For exchange terms only, the Bethe-Salpeter equation can be solved exactly as the bare vertex $\Gamma^{\eta_1\eta_2\eta_1\eta_2}(q_1,q_2+p,q_1+p,q_2)=\Gamma^{\eta_1\eta_2\eta_1\eta_2}(\bs{p})$ only depends on the electronic momentum $\bs{p}$. We can hence assume $\tilde{\Gamma}=\tilde{\Gamma}(p)$ and the sum over $\bs{\tilde{q}}$ in Eq. \eqref{eq:diag7} can be evaluated trivially.

Eq. \eqref{eq:diag7} can then be brought into simple matrix form by defining the vector
\begin{align}
\vec{\Gamma}(\bs{p})\equiv
\begin{pmatrix}
\Gamma^{xxxx}(\bs{p}), &
\Gamma^{xyxy}(\bs{p}), &
\Gamma^{yxyx}(\bs{p}), &
\Gamma^{yyyy}(\bs{p}) \\
\end{pmatrix}\label{eq:diag9}
\end{align}
(and likewise for $\vec{\tilde{\Gamma}}$) and writing $\vec{\Gamma}(\bs{p})=\mathcal{M}(p)\cdot \vec{\tilde{\Gamma}}(p)$ with
\begin{widetext}
\begin{align}
\mathcal{M}(p)=
\begin{pmatrix}
1-\beta N \psi\, \Gamma^{xxxx} & 0 & -\beta N\psi\,\Gamma^{xyxy} & 0\\
0 & 1-\beta N \psi\,\Gamma^{xxxx} & 0 & -\beta N \psi\,\Gamma^{xyxy} \\
-\beta N \psi\,\Gamma^{yxyx} & 0 & 1-\beta N\psi\,\Gamma^{yyyy} & 0 \\
0 & -\beta N\psi\,\Gamma^{yxyx} & 0 & 1-\beta N\psi\,\Gamma^{yyyy} \\
\end{pmatrix}.\label{eq:diag10}
\end{align}
\end{widetext}
Straightforward matrix inversion then yields the effective vertex.

We first shortly discuss how to implement the hard-core constraint beyond the mean field level of chemical potentials. To this end we note that for every $m$-rung real space particle-hole ladder diagram, antiparallel bosonic and fermionic propagator lines corresponds to the same link on the lattice. This is due to ingoing bosonic links matching outgoing fermionic ones and vice versa on every interaction line in exchange processes. As the particle-hole bubble is proportional to the total average dimer density on the corresponding link, the contribution of an $m$-rung process is proportional to the product $\prod_{s=0}^{m} n_D(j_s,n_s)$ over the total dimer densities $n_D=1/4$ of all links $(j_s,n_s)$ occuring in the process. Because we expand around the $t_i=0$ RK-point, we may substitute this product with the classical probability of having all the links occuring in the process occupied, denoted by $Q_c[(j_0,n_0),...,(j_m,n_m)]$.  These correlations can be computed from a Grassmann field theory for the classical dimer problem \cite{samuel1980use}. Attention has to be paid only for interaction lines corresponding to $\tilde{H}_{t_2}$, where the classical probability of having both occuring links occupied would vanish. Instead, we consider the probability of having the two dimers in a relative position which corresponds to the original $t_2$-flip. By this reasoning, the classical dimer correlations can in principle be implemented exactly into our approach. We can achieve a first order approximation by the simple replacement $t_i\rightarrow t_i\cdot 16\,Q_c[(j_s,n_s),(j_{s}+r_{t_i},n_{s}+\Delta n_{t_i})]$, where $(r_{t_i},\Delta n_{t_i})$ correspond to the displacement vector and relative orientation change in a single $t_i$ process. In analogy to \cite{punk2015quantum} this leads to the simple replacements $t_1\rightarrow 4/2\cdot t_1$, $t_2\rightarrow 4/2\cdot t_2$, $t_3\rightarrow 4/\pi\cdot t_3$ that have to be made in all our diagrammatic results presented below.

\begin{figure*}
\centering
\includegraphics[width=\linewidth]{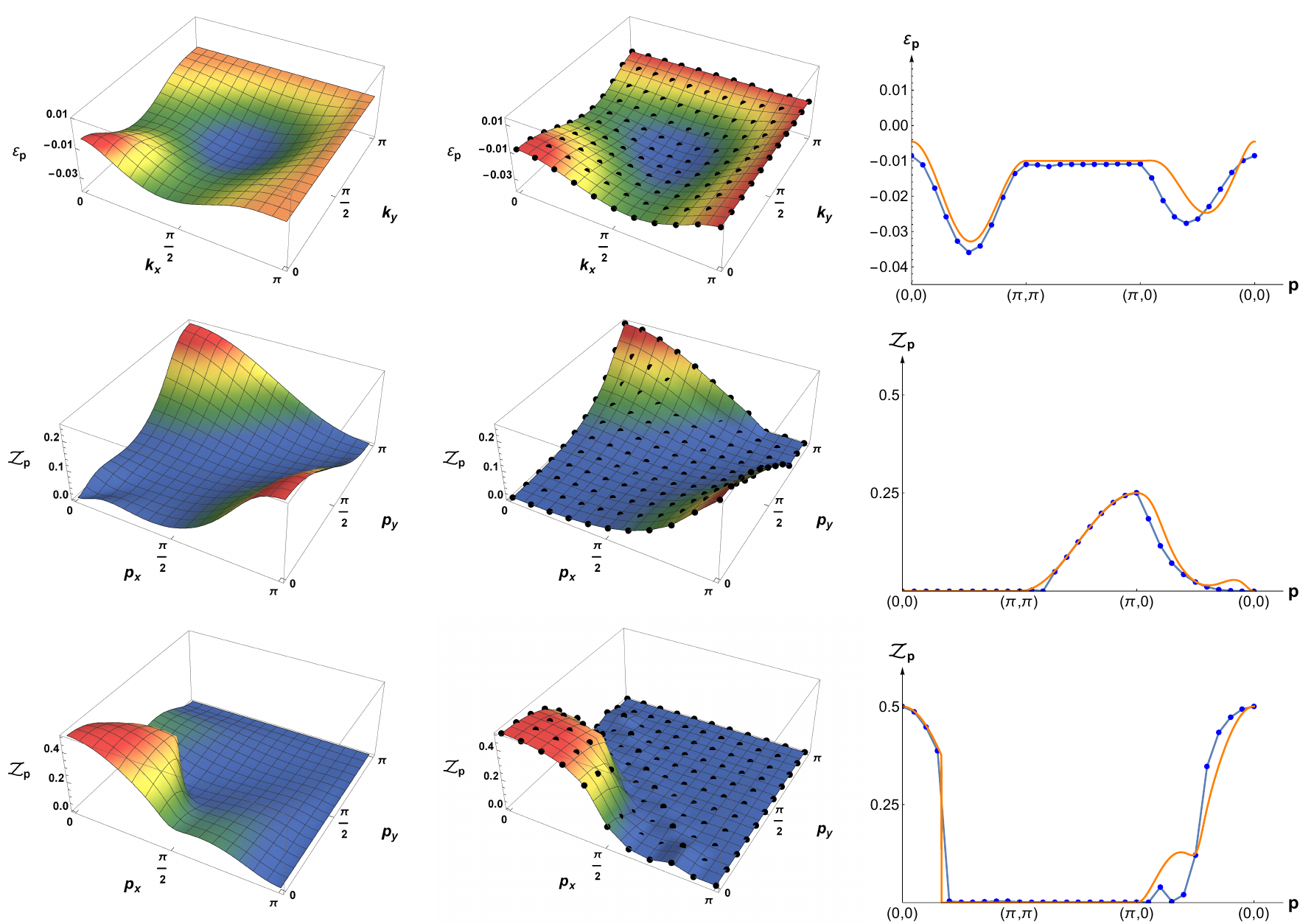}
\caption{Comparison of the dimer dispersion and quasiparticle residuum, computed using the ladder Ansatz and exact diagonalization (ED) for a lattice with $6\times6$ sites. Left column: results from ladder approach; Middle column: results from ED; Right column: line cuts of $\varepsilon_{\bs{p}}$ or $\mathcal{Z}_p$ (Orange: ladder approach; Blue: ED).  Top row: Dispersion for $t_1=-0.01$, $t_2=-0.02$, $t_3=0.01$; Middle row: Residuum for  $t_1=-0.01$, $t_2=-0.02$, $t_3=0.01$; Bottom row: Residuum for $t_1=t_3=0.01$, $t_2=0$.}
\label{fig:residuum1}
\end{figure*}

Using the effective vertex from above and introducing a finite peak width $\gamma$ via $i\omega_p\rightarrow \omega_p+i\gamma$ yields the electronic spectral function
\begin{widetext}
\begin{align}
\mathcal{A}_{\bs{p}}(\omega_p)= 4\gamma\frac{8\mathcal{Z}_{\bs{p}}^0\left( 16\left(\gamma^2+\omega_p^2\right)+K_1(\bs{p})\right)+2K_2(\bs{p})\left(4\omega_p-t_1\cos (p_x)-t_1\cos (p_y)\right)}{\left[16\left(\gamma^2-\omega_p^2\right)+8\omega_pt_1\left(\cos (p_x)+\cos (p_y)\right)+K_1(\bs{p})\right]^2+\left[8\gamma\left(4\omega_p-t_1\cos (p_x)-t_1\cos (p_y)\right)\right]^2},\label{eq:diag11}
\end{align}
\end{widetext}
which assumes the form  of a sum of two  Lorentzians $\mathcal{A}=\mathcal{Z}_{1,\bs{p}}\cdot\frac{2\gamma}{(\omega_p-\omega_1)^2+\gamma^2}+\mathcal{Z}_{2,\bs{p}}\cdot\frac{2\gamma}{(\omega_p-\omega_2)^2+\gamma^2}$. The functions $K_1(\bs{p})$ and $K_2(\bs{p})$ are given in the Appendix while $\mathcal{Z}^0$ is the residuum at $t_i=0$. The peak positions of Eq.(\ref{eq:diag11}) are given by the singularities in the limit $\gamma\rightarrow 0$ and turn out to be 
\begin{align}
\omega_{1,2}=&\frac{1}{4}t_1\left(\cos (p_x)+\cos (p_y)\right) \label{eq:diag12} \\
&\pm\frac{1}{4}\sqrt{t_1^2\left(\cos (p_x)+\cos (p_y)\right)^2+K_1(\bs{p})}. \nonumber
\end{align}
Note that the spectral function \eqref{eq:diag11} only has two sharp peaks and no incoherent background due to the fact that the dimer propagators cannot obtain a dispersion at any order in perturbation theory. Consequently, the particle-hole bubble of a bosonic and a fermionic dimer has a simple pole on the real frequency axis, which gives rise to the simple two-peak structure in the spectral function.

As the insertion of holes corresponds to the removal of the highest band electrons, we can relate the dispersion and residuum for the holes with the rightmost peak of Eq. \eqref{eq:diag11} to obtain
\begin{align}
\varepsilon_{\bs{p}}=&-\omega_1 (\bs{p}) \label{eq:diag13a}
\end{align}
\begin{align}
\mathcal{Z}_{\bs{p}}=\mathcal{Z}_{1,\bs{p}}=&\lim_{\gamma\rightarrow 0}\frac{1}{2}\gamma\mathcal{A}_{\bs{p}}(\omega_1(\bs{p})) \label{eq:diag13b}
\end{align}
for hole-dispersion and -residuum. Examples for different values of $t_i$ are compared to ED results in Fig. \ref{fig:residuum1}. For the chosen parameter regime $t_i\ll J$ the assumption of an RK-background is valid and yields good agreement with ED results. This includes parameter sets with non-vanishing $t_2$-interactions, which were treated via the effective exchange interaction of Eq. \eqref{eq:diag14}. We note that the hole dispersion of Eq. \eqref{eq:diag13a} reproduces exactly the dispersion obtained in \cite{punk2015quantum} by a perturbative Ansatz. Our approach can hence be viewed as an extension of this Ansatz which additionally yields the quasiparticle residuum.

\section{Discussion and Conclusions}\label{M_S5}

Our numerical results show that the dimer model introduced in Ref.~\cite{punk2015quantum} has a sizeable pseudogap in the antinodal region of the Brillouin zone close to $\mathbf{k}=(0,\pi)$ and symmetry related momenta. Moreover, it's momentum dependence clearly deviates from a simple d-wave form, in accordance with observations in the pseudogap regime of underdoped cuprates. 

It is important to emphasize here that we always fine-tune our model to the RK-point at $J=V$, where the ground-state of the undoped model is a $U(1)$ spin liquid. Away from this point the $U(1)$ spin liquid is confining and unstable towards symmetry broken valence bond solid states. Upon doping away from the RK-point the dimer model \eqref{HAM} thus realizes a $U(1)$-FL*, which again is expected to be confining at long length-scales. This problem can be circumvented by allowing for diagonal dimers between next-nearest neighbor sites as well. In this case the undoped RK-model has a stable $Z_2$ spin liquid ground-state in an extended parameter regime \cite{moessner2001resonating}. Accordingly, upon doping the ground-state of the appropriately modified model in Eq.~\eqref{HAM} is expected to be a $Z_2$-FL*, which is a stable phase of matter \cite{Patel2016}. Nevertheless, we don't expect a qualitatively different behavior of the single-electron spectral function when including diagonal dimers. In particular, the pseudogap in the antinodal region is expected to be a robust property of the dimer model. We leave a detailed analysis of this problem open for further study.

It's also worthwile to contrast our results with numerical dynamical cluster approximation (DCA) studies of the Hubbard model on the square lattice, where a sizeable pseudogap in the spectral function at the antinode was found as well \cite{gull2010momentum,gull2013superconductivity}. We point out here that two site cluster dynamical mean field theory (DMFT) studies of the Hubbard model indeed showed that the electron configuration on the two site cluster at low dopings is dominated by the same two states that are used here to span the Hilbert space of the dimer model on the full lattice, i.e.~the bosonic and the fermionic dimer \cite{ferrero2009pseudogap}. This suggests that the dimer model introduced in Ref.~\cite{punk2015quantum} provides an effective low energy description of the square lattice Hubbard model at large $U$ and low doping.

\acknowledgements

We thank D.~Pimenov and F.~Dorfner for valuable discussions and S.~Sachdev for comments on the manuscript. This research was supported by the German Excellence Initiative via the Nanosystems Initiative Munich (NIM).

\appendix

\section{Diagrammatic Approach}

We provide some additional information on the ladder approach of Sec.(\ref{DIAG}). The bare vertex $\Gamma (\bs{p})$ from the Bethe-Salpeter Eq. \eqref{eq:diag7} is determined from the momentum space form of the dimer Hamiltonian $H_1$, including $\tilde{H}_{t_2}$. The relevant bare vertices in Eq. \eqref{eq:diag9} and \eqref{eq:diag10} are
\begin{align}
\Gamma^{xxxx}(p)=&\frac{2t_1}{\beta N}\cos(p_y), \label{eq:Adiag1}\\
\Gamma^{yyyy}(p)=&\frac{2t_1}{\beta N}\cos(p_x), \\
\Gamma^{xyxy}(p)\equiv &C(\bs{p})= \\
=&\frac{t_3}{\beta N}\bigl( e^{i(p_x+p_y)}+e^{ip_x}+e^{i(p_y-2p_x)}+e^{-2ip_x} \nonumber\\
&+e^{2ip_y}+e^{i(2p_y-p_x)}+e^{-ip_y}+e^{-i(p_x+p_y)}\bigr)+ \nonumber\\
&+\frac{t_2}{\beta N}\bigl(1+e^{ip_x}\bigr)\bigl(1+e^{-ip_y}\bigr), \nonumber\\ 
\Gamma^{yxyx}(p)=&C^*(\bs{p})=C(-\bs{p}).
\end{align}
The functions $K_1(\bs{p})$ and $K_2(\bs{p})$ which contribute to the spectral function from Eq. \eqref{eq:diag11} are given by
\begin{align}
&K_1(\bs{p})=t_3^2\biggl[8+4\cos (p_x)+4\cos (3p_x)+2\cos (p_x-3p_y)+ \label{eq:Adiag2}\\
&+4\cos (p_x-2p_y)+4\cos (2p_x-2p_y)+4\cos (p_x-p_y)+ \nonumber\\
&+4\cos (2p_x-p_y)+2\cos (3p_x-p_y)+4\cos (p_y)+ \nonumber\\
&+4\cos (3p_y)+4\cos (p_x+p_y)+4\cos (2p_x+2p_y)+ \nonumber\\
&+4\cos (2p_x+p_y)+2\cos (3p_x+p_y)+ \nonumber\\
&+4\cos (p_x+2p_y)+2\cos (p_x+3p_y)\biggr] \nonumber\\
&-t_1^2\biggl[2\cos (p_x-p_y)+2\cos (p_x+p_y)\biggr]+ \nonumber
\end{align}
\begin{align}
&+t_2^2\biggl[4+4\cos (p_x)+4\cos (p_y)+ \nonumber\\
&+2\cos (p_x-p_y)+2\cos (p_x+p_y)\biggr]+ \nonumber\\
&+t_2t_3\biggl[4+6\cos (p_x)+6\cos (p_y)+4\cos (2p_x)+ \nonumber\\
&+4\cos (2p_y)+2\cos (3p_x)+2\cos (3p_y)+4\cos (p_x-3p_y)+ \nonumber\\
&+4\cos (3p_x-p_y)+2\cos (2p_x-3p_y)+2\cos (3p_x-2p_y)+\nonumber\\
&4\cos (p_x-2p_y)+4\cos (2p_x-p_y)+4\cos (2p_x-2p_y)+ \nonumber\\
&+8\cos (p_x+p_y)+2\cos (2p_x+p_y)+2\cos (p_x+2p_y)\biggr], \nonumber\\
&K_2(\bs{p})=t_3\biggl[2+3\cos (p_x)+ 2\cos (2p_x)+\cos (3p_x)+ \nonumber\\
&+2\cos (3p_x-p_y)+\cos (2p_x-3p_y)+2\cos (p_x-2p_y)+ \nonumber\\
&+2\cos (2p_x-2p_y)+\cos (3p_x-2p_y)+2\cos (2p_x-p_y)+ \nonumber\\
&+2\cos (3p_x-p_y)+3\cos (p_y)+2\cos (2p_y)+\cos (3p_y)+ \nonumber\\
&+4\cos (p_x+p_y)+\cos (2p_x+p_y)+\cos (p_x+2p_y)\biggr] \nonumber\\
&-t_1\biggl[2\cos (p_x)+\cos (2p_x)+2\cos (p_y)+\cos (2p_y)\biggr]+ \nonumber\\
&+t_2\biggl[4+4\cos (p_x)+4\cos (p_y)+ \nonumber\\
&+2\cos (p_x-p_y)+2\cos (p_x+p_y)\biggr]. \nonumber
\end{align}
Note that these functions satisfy $K_{1/2}(-\bs{p})=K_{1/2}(\bs{p})$ and $K_1$ is even in the parameters $t_i$ while $K_2$ is odd. We use the notation $K_{1,\{-t_i\}}=K_{1,\{t_i\}}$, $K_{2,\{-t_i\}}=-K_{2,\{t_i\}}$. From these properties and the expression Eq. \eqref{eq:diag11} for the spectral function we can easily deduce
\begin{align}
\mathcal{A}_{-\bs{p}}(\omega_p)=&\mathcal{A}_{\bs{p}}(\omega_p), \label{Adiag4}\\
\mathcal{A}_{\{-t_i\},\bs{p}}(\omega_p)=&\mathcal{A}_{\{t_i\},\bs{p}}(-\omega_p), \nonumber\\
\varepsilon_{\{a\cdot t_i\},\bs{p}}=&a\cdot \varepsilon_{\{t_i\},\bs{p}}, \nonumber
\end{align}
\begin{align}
\mathcal{Z}_{\{a\cdot t_i\},\bs{p}}=&\mathcal{Z}_{\{t_i\},\bs{p}}. \nonumber
\end{align}
for any positive $a$. Using the second of these equations one can prove by a straightforward calculation the interesting relation
\begin{align}
\mathcal{Z}_{\{-t_i\},\bs{p}}&+\mathcal{Z}_{\{t_i\},\bs{p}}=\\
=&\lim_{\gamma\rightarrow 0}\frac{1}{2}\gamma\left[\mathcal{A}_{\{t_i\},\bs{p}}(\omega_1(\bs{p}))+\mathcal{A}_{\{t_i\},\bs{p}}(\omega_2(\bs{p}))\right]= \nonumber\\
=& \mathcal{Z}_{\bs{p}}^0=\frac{1}{4}\left[\cos^2 (p_x/2)+\cos^2 (p_y/2)\right]. \nonumber
\end{align}

\bibliography{Dimer}
\bibliographystyle{apsrev4-1}

\end{document}